# Theory of Periodic Sequence:
# Bridging Time-Domain and Frequency-Domain for Computational Electromagnetics

Ben You, Ke Wu*

*aPoly-Grames Research Center, Department of Electrical Engineering, University of Montréal, Montréal, H3T 1J4, Québec, Canada*

**Abstract**

Time-periodic form or expression is a ubiquitous natural and man-made phenomenon observable in all the scientific and engineering disciplines. In this article, we propose a theory of periodic sequence (TPS), which can be formulated as a foundational theory for computational sciences and engineering, to transform arbitrary time-periodic electromagnetic (EM) problems into a computational space with mapped discrete events, which is characterized in neither frequency domain nor time domain. Within the TPS framework, periodic-sequential Maxwell's curl equations are decomposed and decoupled to independent and paralleled instances via designated mappings. The fundamental solutions and mapped responses of EM periodic sequences are elucidated, and corroborated by RF/microwave measurements. The nature of outstanding computational parallelism and the unique frequency-independent property make TPS a promising methodology for computational electromagnetics such as analysis of high-speed signal integrity and broadband RF transmission.

1. **Introduction**

Time-periodic phenomena can be observed ubiquitously in the natural world and the engineering field. In fact, such a time-scale periodicity or periodic sequence that inherently characterizes parametric motions, dynamics, and evolutions, presents fundamental features and phenomena in the disciplines of physics and others including electrodynamics [1, 2], acoustics [3, 4, 5], thermodynamics [6, 7, 8], and quantum mechanics [9, 10, 11], etc.. With the fast development of modern computers and computing techniques, any natural and engineered dynamics such as the propagation of waves can be easily described and explained by temporal-spatial sequences in terms of Fourier decompositions or compositions. In this connection, it inspires and guides us to utilize specific periodic sequences of choice for developing theories of physics, especially for wave equations under time-periodic boundary conditions (TPBC). The Maxwell equations-based electromagnetic (EM) theories [12, 13, 14, 15, 16] have always been formulated on either sinusoids (frequency domain) or transient pulses (time domain), corresponding to single tone (time-harmonic) or continuous spectrum in terms of frequency representations. In fact, they can be regarded as the special cases of temporal-periodic wave sequences, the majority of which are barely studied—those have a finitely-countable spectrum in the frequency domain, carrying valuable waveform and modulated signal information and possessing parallel features for creating a ultra-fast computational EM platform.

In this manuscript, we report a unique and powerful computational methodology, referred to as theory of periodic sequence (TPS), to depict a state-steady process of time-periodic electromagnetic (EM) fields and waves. The Maxwell's curl equations are completely sequentialized and further decoupled in a transform domain by designated mappings. The corresponding parametric representations and fundamentals of periodic sequence-based electromagnetic dynamics, including lossless/lossy scenario, energy conservation, numerical dispersion, and accuracy consideration, are elaborated and validated by experiments. We will reveal and elucidate the interesting and inherent properties of TPS. For example, the computational scale is independent of frequency, which provides an unprecedented flexibility of creating paralleled computational algorithms. We will demonstrate the promising potential of TPS by analyzing a broadband EM coupling problem, providing a glimpse into its superior computational efficiency and high accuracy.

2. **Fundamentals**

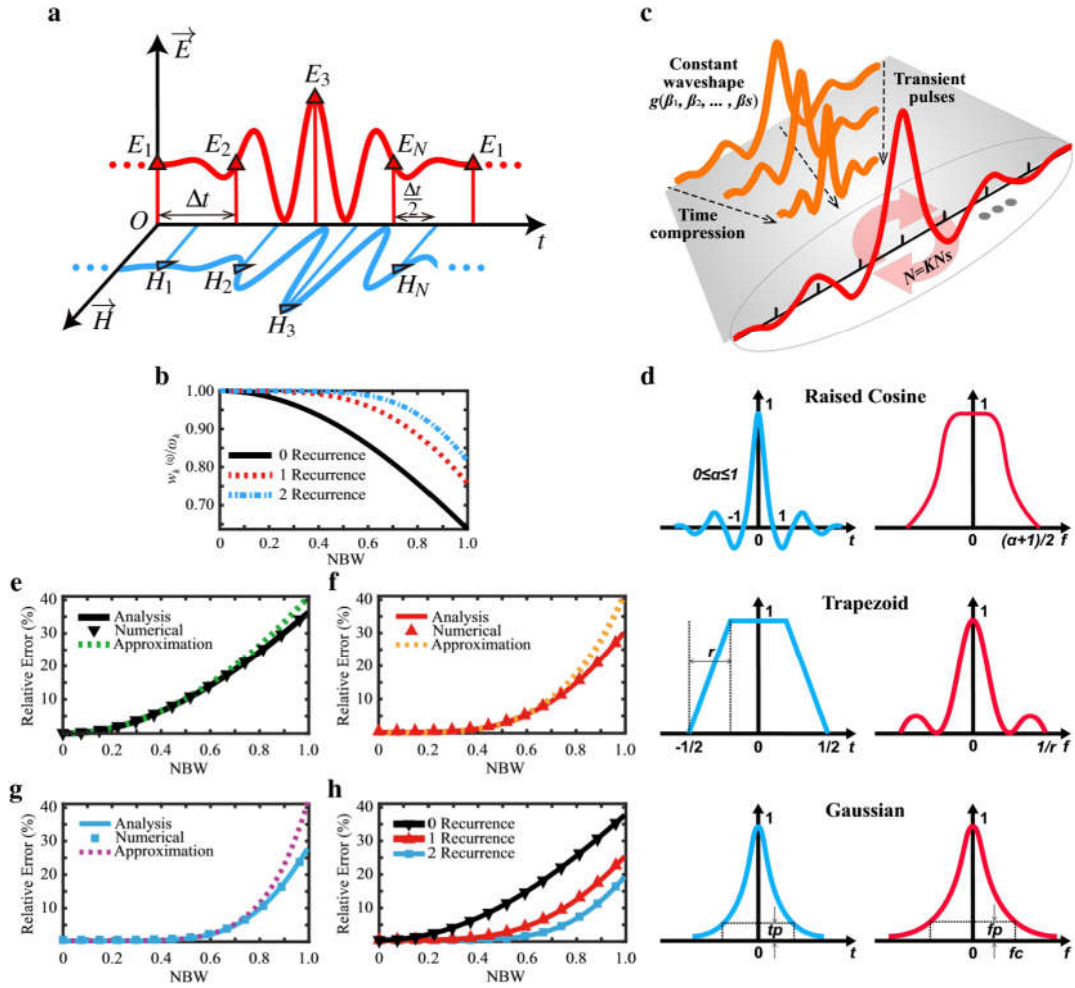

Figure 1: **Basic theory of EM periodic sequences. a** Full time-discretization of Maxwell' curl equations under TPBCs. To simulate discrete EM events, the sampling instant of magnetic field is time-shifted with respect to that of electric field with half-step delay ($\Delta t/2$). **b** Normalized quantized spectrum $w_k^{(q)}/\omega_k$ as a function of NBW for different orders of recurrent difference schemes. **c** Generational process of periodic sequential excitation. **d** Waveform parameter and cut-off criterion of commonly used periodic sequences corresponding to the Supplementary Eqn.(50)-(52). Raised cosine (first row): α represents the roll-off coefficient. Trapezoid (second row): $r$ denotes the rise-time ratio. Gaussian (third row): $tp$ and $fp$ are the truncation percentage of time waveform and frequency spectrum, respectively. **e-g** Relative error of propagation constant corresponding to the main mode of WR-284 (72.14mm×34.04mm): 0-2 recurrence. 'Analysis' curve is derived by Supplementary Eqn.(61) while 'Numerical' curve is based on 2D finite element method (FEM). 'Approximation' corresponds to Supplementary Eqn.(62)-(64). **h** Numerical dispersion vs recurrent order of difference scheme.

In simple and Ohm-lossless media, the Maxwell's curl equations under TPBCs are written as follows,

$$\begin{cases} \nabla \times \vec{E}(t) = -\mu \dfrac{\partial \vec{H}(t)}{\partial t} \\ \nabla \times \vec{H}(t) = +\epsilon \dfrac{\partial \vec{E}(t)}{\partial t} \end{cases} \quad (1)$$

where $\vec{E}(t) = \vec{E}(t+T)$ and $\vec{H}(t) = \vec{H}(t+T)$. For simplicity, all the spatial variables are dropped out of (1). To achieve a full discretization of the temporal components in (1), all field variables are sampled equidistantly in the time domain by a leapfrog scheme with time step $\Delta t$, as depicted in Fig.1a; while their time-partial derivatives are approximated by the first-order difference scheme:

$$[D_t] = \begin{bmatrix} -1 & 0 & 0 & \cdots & 1 \\ 1 & -1 & 0 & \cdots & 0 \\ 0 & 1 & -1 & \cdots & 0 \\ \vdots & \vdots & \ddots & \ddots & \vdots \\ 0 & 0 & \cdots & 1 & -1 \end{bmatrix}_{N \times N} \quad (2)$$

where sequential period is denoted by $N = T/\Delta t$. $[D_t]$ is a circulant matrix, of which the fringe elements are set to characterize the TPBCs. Subsequently, the periodic-sequential curl equations can be expressed as

$$\begin{cases} \nabla \times \vec{E}[n] = \dfrac{\mu}{\Delta t}[D_t]\vec{H}[n] \\ \nabla \times \vec{H}[n] = \dfrac{\epsilon}{\Delta t}[D_t]^+\vec{E}[n] \end{cases} \quad (3)$$

In which superscript + represents the conjugate transpose. According to the leapfrog scheme of EM fields, we can introduce two unitary matrices (see Supplementary Note 1):

$$[T^{e,h}] = \left( \dfrac{SGN^{n_{e,h}}(k)}{\sqrt{N}} W_N^{-(k-1)\left(n - \frac{n_{e,h}}{2}\right)} \right)_{nk} \quad (4)$$

where $n_{e,h} = 2,1$, $W_N = exp(i\,2\pi/N)$ and $SGN(k)$ is given by

$$SGN(k) = \begin{cases} +1, & k = 1,2,\cdots,\lceil \dfrac{N+1}{2} \rceil \\ -1, & k = \lceil \dfrac{N+1}{2} \rceil + 1, \cdots, N \end{cases} \quad (5)$$

$\lceil \cdot \rceil$ denotes the ceiling truncation. (4) implies that there exists a new transform domain, which is neither frequency domain nor time domain, for representing any EM fields based on the periodic sequences in (3). To distinguish it from the canonical frequency domain, we define it as 'w domain', in which the fields conserve the property of conjugate symmetry and (3) can then be transformed and decoupled to a group of independent equations, referred to as w-domain curl equations (see Supplementary Note 2 for detailed derivation):

$$\begin{cases} \nabla \times \vec{e}_k = +iw_k\mu\,\vec{h}_k \\ \nabla \times \vec{h}_k = -iw_k\epsilon\,\vec{e}_k \end{cases} \quad (6)$$

where

$$w_k = \dfrac{2}{\Delta t}\sin\left[\dfrac{\pi}{N}(k-1)\right]SGN(k) \quad (7)$$

$w_k$ possesses the conjugate anti-symmetric property, i.e., $w_k = -w_{N+2-k}$. This anti-symmetry leads to real sequential response after inversely mapping the solutions of (6) to the original discrete-time domain. For the Ohm-lossy scenario, the derivations can be obtained in a similar way (see Supplementary Note 3 for details). We must stress that the EM periodic sequences still and always obey the law of energy conservation in the w domain (see Supplementary Note 4 for details).

To avoid aliasing, we stipulate that the sampling rate should satisfy $f_s =\geq 2(k_{max}-1)/T$ where $k_{max}$ is the maximum index of the nonzero w-domain components corresponding to the given periodic excitation. For the sake of comparison, we define the normalized bandwidth (NBW) as $NBW = 2(k_{max}-1)/N$. Note that the approximation scheme of time-partial derivative directly decides the numerical accuracy.

Although the first-order difference scheme (2) provides a concise analysis for developing the fundamentals of TPS, it is associated with collateral numerical losses. Therefore, for practical applications, we adopt a recurrent difference scheme that offers higher accuracy. To be more specific, the recurrent form of (7) can be expressed by (see Supplementary Note 5 for derivation process)

$$w_k^{(q)} = \frac{w_k^{(0)}}{Sa(w_k^{(q-1)}/2)} \qquad (8)$$

where $w_k^{(0)} = \omega_k = 2\pi f_k = 2\pi(k-1)/T$. Given $q \to \infty$, $w_k^{(q)} \to \omega_k$. Fig.1b shows the normalized $w_k^{(q)}$ as functions of different recurrent orders. Obviously, the numerical error would rapidly decrease with the rising of order.

Unlike traditional harmonic analysis, the periodic sequential excitation plays an important role in TPS since it carries waveform and signal information. Fig.1c depicts a general building process of the EM periodic sequence, which origins from transient pulses with arbitrary waveform. Note that transient pulses with different time span but identical waveform are corresponding to the same periodic sequence. We stipulate that $N$ can be equidistantly partitioned to $K$ segments, i.e., $N = KN_s$ where $N_s$ denotes the code duration or unit interval (UI). The maximum index of non-zero components can be expressed as (see Supplementary Note 6 for detailed derivations)

$$k_{max} = \lfloor g(\beta_1, \beta_2, \cdots, \beta_S)K \rfloor \qquad (9)$$

Here, $\lfloor \cdot \rfloor$ represents the floor truncation, function $g(\cdot)$ is irrelevant to frequency and only decided by waveform parameters and cut-off criterion of the original transient pulse, as illustrated by Fig.1d. Given a fixed UI-measured period $K$, we can arbitrarily compress the original transient pulse while still preserving the same computational scale in the w domain. From the perspective of numerical dispersion, Fig.1e-g illustrate the relative error of a propagation constant (Supplementary Eqn.(56)) of a rectangular waveguide (RWG), corresponding to different recurrent orders. The relative error of the main mode $TE_{10k}$ is proportional to NBW to the power 2, 4 and 6, respectively (see Supplementary Note 7 for details). Fig.1h indicates that the numerical dispersion can be suppressed to a satisfactory level after a few recurrences of (8).

3. **Experimental Validation**

First, we have investigated the propagation of a modulated-Gaussian periodic sequence in standard RWGs. As excitation, those periodic sequences possess an identical waveform but different quantized spectrum, as illustrated in Fig.2a-b. The deviation of quantized spectrum reflects the numerical error caused by difference schemes and will be enlarged with the increase of index $k$. Fig.2c shows the corresponding output response of RWG. It is clear to see that in the case of a large NBW, the first-order difference scheme results in a significant numerical dispersion. Hence, to ensure a good numerical accuracy, we adopt the second order recurrent difference scheme in practical applications. To validate our proposed theory, we have conducted an experiment of validation as shown in Fig.2d. Subject to the maximum sampling rate of our arbitrary waveform generator (AWG), a standard RWG (WR-284) is taken as the testing device, which is excited by a modulated-Gaussian periodic sequence, as depicted in Fig.2e-f. Fig.2g illustrates the output periodic responses of the RWG corresponding to the proposed method and the oscilloscope (OSC) display, respectively. It is clearly to see that the theoretical curve is matching well with the experimental counterpart. A slight distortion observed here is mainly attributed to the non-ideal performance of the AWG. Nevertheless, the measured results have substantiated and validated the correctness of our proposed theory.

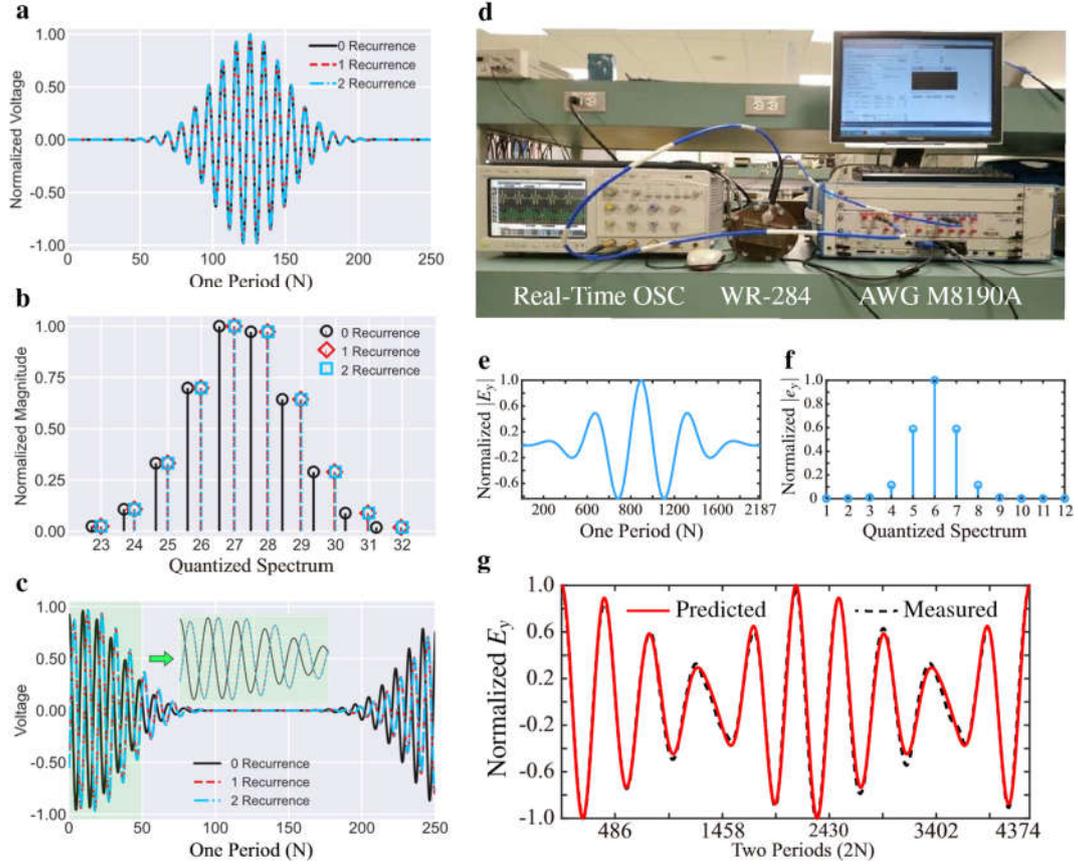

Figure 2: **Propagation of modulated-Gaussian periodic sequence in RWG. a** Input modulated-Gaussian periodic sequence vs recurrent orders of difference scheme. The bandwidth of the baseband Gaussian equals to 17.5GHz and the carrier frequency is set to 97.5GHz. $(tp, fp)$, $N_s$ and $K$ are set to (0.01,0.01), 50 and 5, respectively. **b** Corresponding quantized spectrum. **c** Periodic sequential responses of RWG WR10 (2.54mm×1.27mm×50mm). **d** The experimental system for verification and validation. It is composed of an AWG Keysight M8190A (maximum sampling rate is 12 Ga/s), the under-tested WR-284 (2.14mm×34.04mm×420mm) and a real-time Oscilloscope (OSC) Agilent Infiniium DSO81204B (maximum sampling rate is 40Ga/s). **e-f** The input modulated-Gaussian periodic sequence and its quantized spectrum. $(tp, fp) = (0.01, 0.1)$, $N_s = 729$ and $K = 3$. The corresponding data rate and carrier frequency equal to 1.5GHz and 2.5GHz, respectively. **g** Output periodic response of WR-284: theoretical prediction vs OSC measurement. The time delay caused by cables and fixtures has been removed.

Moreover, we extend the concept of scattering parameters [17] to the w domain so as to derive a periodic response in a closed region scenario (See Supplementary Note 8 for details). To investigate the coupling effect between two synchronized EM periodic sequences (generated by broadband pulses), we have simulated and fabricated four crosstalk boards with different routine delays (denoted by case 1-4), as illustrated in Fig.3d-g. The experimental system is shown in Fig.3a. To provide a reasonable comparison between simulation and measurement, we send a raised-cosine (α = 1) broadband pulse (up to 45GHz) to AWG, and take the average of output (4096 waveforms) as the excitation. The discrepancy between the two synchronized AWG outputs and the ideal reference can be easily observed from Fig.3b-c. Since $k_{max} = 10$ corresponds to a small computational scale, we implement one-batch multiprocessing to solve those components simultaneously. To provide a quantitative comparison, we introduce the Kullback-Leibler (KL) divergence as the figure of merit. In mathematics, the KL divergence is a measure of deviation between two nonnegative vectors[18]:

$$D_{KL}(u,v) = \sum_{n=1}^{N}\left\{u_n \ln \frac{u_n}{v_n} - u_n + v_n\right\} \tag{10}$$

where $u_n = |\widetilde{q_1}[n]|/|\widetilde{q_2}[n]|_2 + \varepsilon$, $v_n = |\widetilde{q_2}[n]|/|\widetilde{q_2}[n]|_2 + \varepsilon$ and $\varepsilon = 10^{-12}$.

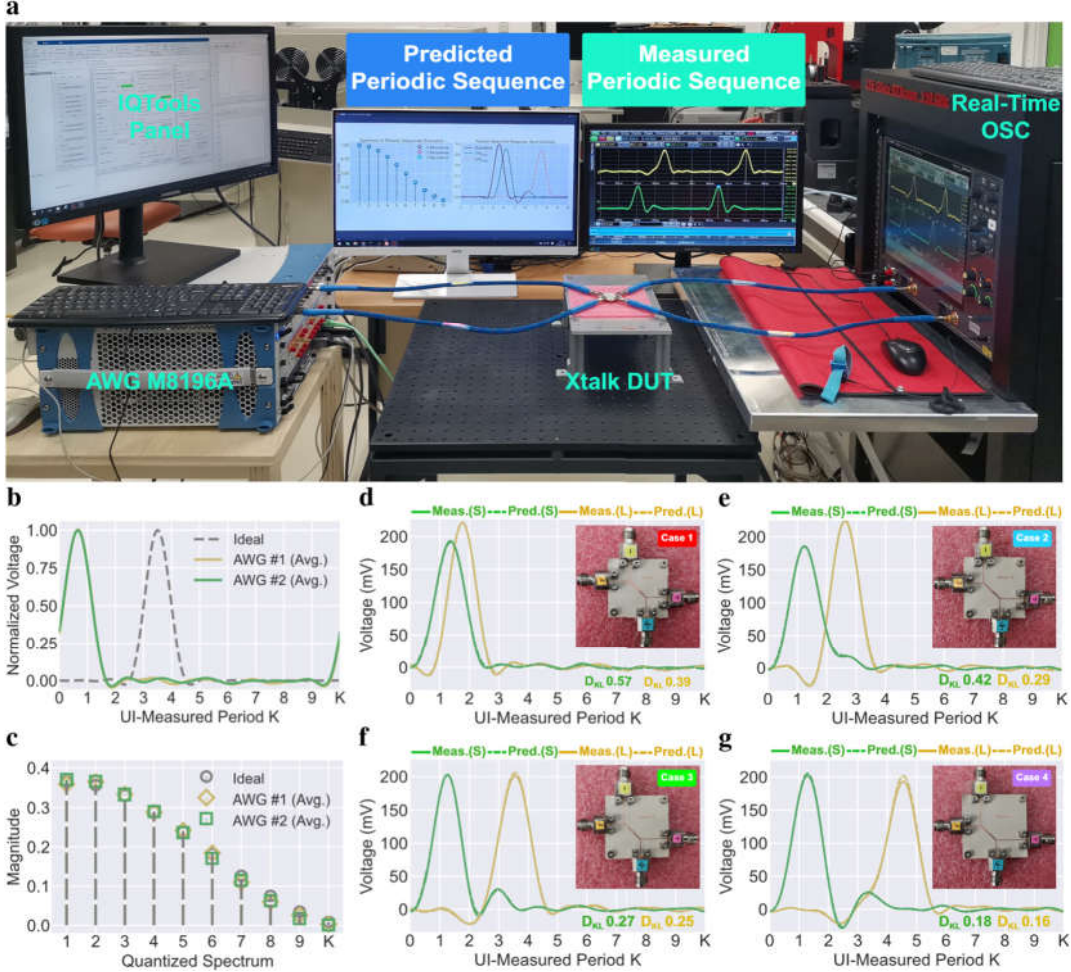

Figure 3: **Coupling between EM periodic sequences according to 45GHz-broadband pulse. a** Measurement set-up. The system is composed of an AWG Keysight M8196A (maximum sampling rate of 92GSa/s), the under-tested crosstalk boards, a real-time OSC UXR0702AP (maximum sampling rate of 256GSa/s) and four V-band cables. The outputs of AWG are controlled by the panel of Keysight IQtools. To avoid aliasing, the sampling rate of AWG is set to 90GSa/s. After the in-system calibration, the AWG channels are synchronized (the skew is less than 0.2ps), and the attenuation is considerably alleviated. The loss, delay and inter-channel skew of the V-band cables are removed by the common procedure of the insystem calibration and the Keysight D9020ASIA advanced SI software. The under-tested microstrip boards are made of Rogers RO3003 substrate (5mil thickness), operating from DC to 50GHz. Both routines (width 0.3mm) are terminated by V-band end launchers. To provide significant crosstalk effects, the coupling length and gap spacing are set to 10mm and 0.2mm, respectively. The difference of length between routine port1-port2 ('long') and routine port3-port4 ('short') is equal to 1/3/5/7mm. **b, c** Input periodic raised-cosine sequences and the corresponding spectrum—ideal (gray dashed), the outputs of AWG channel 1 (golden solid) and channel 2 (green solid), as the average results of consecutive 4096 waveforms. **d-g** Average periodic responses of the under-tested circuits (case 1-4): algorithm predictions vs measured results.

Note that the KL divergence is non-negative and equal to zero if and only if two sequences are identical, i.e. $u = v$. Fig.3d-g illustrate the periodic responses of the under-tested circuits (case 1-4) according to the prediction and the measurement, respectively. In each case, the predicted curves are fitting well with the measured ones, and this observation is further corroborated by the corresponding small KL divergences.

## 4. Conclusion

In conclusion, we have presented, to our best knowledge, the first open publication and development on TPS and its applications in connection with time-periodic EM waves. TPS can successfully depict the propagation process of EM periodic sequences with arbitrary waveform. The validation experiments indicate a good consistency between theoretical predictions and measured results. The frequency-independent property of periodic sequences and the parallel nature of the w-domain curl equations makes TPS a promising tool in analysis of high-speed signal integrity (SI) systems and modelling of (ultra-)wideband wired/wireless channels.

**Supplementary Note 1: Selection of Transform Operators**

As mentioned in our manuscript, the periodic sequential Maxwell's equations can be expressed as

$$\begin{cases} \nabla \times \vec{E}[n] = \dfrac{\mu}{\Delta t}[D_t]\vec{H}[n] \\ \nabla \times \vec{H}[n] = \dfrac{\epsilon}{\Delta t}[D_t]^+\vec{E}[n] \end{cases} \tag{11}$$

where

$$[D_t] = \begin{bmatrix} -1 & 0 & 0 & \cdots & 1 \\ 1 & -1 & 0 & \cdots & 0 \\ 0 & 1 & -1 & \cdots & 0 \\ \vdots & \vdots & \ddots & \ddots & \vdots \\ 0 & 0 & \cdots & 1 & -1 \end{bmatrix}_{N \times N} \tag{12}$$

Based on (11), the wave equation can be easily derived as

$$\nabla \times \nabla \times \vec{E}[n] - \dfrac{\epsilon\mu}{\Delta t^2}[D_{tt}]\vec{E}[n] = 0 \tag{13}$$

where

$$[D_{tt}] = [D_t][D_t]^+ = \begin{bmatrix} 2 & -1 & 0 & \cdots & 1 \\ -1 & 2 & -1 & \cdots & 0 \\ 0 & -1 & 2 & \cdots & 0 \\ \vdots & \vdots & \ddots & \ddots & \vdots \\ -1 & 0 & \cdots & -1 & 2 \end{bmatrix}_{N \times N} \tag{14}$$

$[D_{tt}]$ denotes the second-order difference scheme and also possesses circulant elements. In mathematics, circulant matrices are categorized to be of 'normal' type—they can always be diagonalized by unitary matrices. Note that those transform operators should be subject to the following rules:

- The transform operators are able to diagonalize both the first-order (12) and second-order (14) difference schemes.
- After applying inverse mapping, the solution of wave equation (13) should yield a real number at any observation point.

Consequentially, for electric field, we define the bilateral mapping as

$$[T^e] = \left(\dfrac{1}{\sqrt{N}} W_N^{-(k-1)(n-1)}\right)_{nk} \tag{15}$$

Accordingly, for magnetic field, we have

$$[T^h] = \left(\dfrac{1}{\sqrt{N}} SGN(k) W_N^{-(k-1)\left(n-\frac{1}{2}\right)}\right)_{nk} \tag{16}$$

Where $W_N = exp(i\,2\pi/N)$ and $SGN(k)$ is given by

$$SGN(k) = \begin{cases} +1, & k = 1, 2, \cdots, \lceil\dfrac{N+1}{2}\rceil \\ -1, & k = \lceil\dfrac{N+1}{2}\rceil + 1, \cdots, N \end{cases} \tag{17}$$

$\lceil \cdot \rceil$ stands for the ceiling truncation.

The w-domain electric field and magnetic fields can be respectively written as

$$\vec{e}[k] = [T^e]^+ \vec{E}[n] = \frac{1}{\sqrt{N}} \sum_{n=1}^{N} \vec{E}[n] W_N^{(k-1)(n-1)} \tag{18}$$

and

$$\vec{h}[k] = [T^h]^+ \vec{H}[n] = \frac{1}{\sqrt{N}} \sum_{n=1}^{N} \vec{H}[n] SGN(k) W_N^{(k-1)\left(n-\frac{1}{2}\right)} \tag{19}$$

(18)/(19) is conjugate symmetric if $\vec{E}[n]/\vec{H}[n]$ is real. To be more specific, for each index $k$, we have

$$\begin{cases} \vec{e}_{N+2-k} = \vec{e}_k^* \\ \vec{h}_{N+2-k} = \vec{h}_k^* \end{cases} \tag{20}$$

(17) imposes the identical symmetry on both electric field and magnetic field, and it is a compulsory condition, otherwise, the w-domain magnetic field will be conjugate anti-symmetric (due to the shift or delay $\Delta t/2$) and incur non-physical effect which violates the law of energy conservation.

## Supplementary Note 2: w-domain Maxwell's Curl Equations

Based on (18) and (19), the first- and second-order difference matrices can be diagonalized as

$$= [T^e]^+ [D_t][T^h] = 2i \sin\left[\frac{\pi}{N}(k-1)\right] SGN(k) \tag{21}$$

and

$$[\Lambda_k]^2 = [T^e]^+ [D_{tt}][T^e] = -4\sin^2\left[\frac{\pi}{N}(k-1)\right] \tag{22}$$

, respectively. (22) implies that $[D_{tt}]$ has $[(N+1)/2]$ non-repetitive eigenvalues due to the symmetric property:

$$\lambda_k = \lambda_{N+2-k}, \quad k = 1,2,\cdots, \lceil\frac{N+1}{2}\rceil \tag{23}$$

In turn, the w-domain curl equations can be formulated by

$$\begin{cases} \nabla \times \vec{e}_k = +i w_k \mu\, \vec{h}_k \\ \nabla \times \vec{h}_k = -i w_k \epsilon\, \vec{e}_k \end{cases} \tag{24}$$

where

$$w_k = \frac{2}{\Delta t} \sin\left[\frac{\pi}{N}(k-1)\right] SGN(k) \tag{25}$$

Note that $w_k$ preserves the conjugate anti-symmetric property, i.e., $w_k = -w_{N+2-k}$. Substituting $k = N+2-k$ to (24), we have

$$\begin{cases} \nabla \times \vec{e}_{N+2-k} = +iw_{N+2-k} \; \mu \; \vec{h}_{N+2-k} \\ \nabla \times \vec{h}_{N+2-k} = -iw_{N+2-k} \; \epsilon \; \vec{e}_{N+2-k} \end{cases}$$
$$\Downarrow$$
$$\begin{cases} \nabla \times \vec{e}^{\,*}_{N+2-k} = +iw_k \mu \; \vec{h}^{\,*}_{N+2-k} \\ \nabla \times \vec{h}^{\,*}_{N+2-k} = -iw_k \epsilon \; \vec{e}^{\,*}_{N+2-k} \end{cases} \tag{26}$$

Comparing (26) with (24), it is obvious to see that the anti-symmetric property of $w_k$ ensures the conjugate symmetry of the solution of w-domain curl equations. Once mapping the fields back to the original domain, they stay real as required.

## Supplementary Note 3: Ohm-Loss Scenario

In Ohm-loss scenario, the periodic-sequential curl equations can be written as

$$\begin{cases} \nabla \times [\overrightarrow{E_n}] = \dfrac{\mu}{\Delta t}[D_t][\overrightarrow{H_n}] \\ \nabla \times [\overrightarrow{H_n}] = \left(\sigma[O] + \dfrac{\epsilon}{\Delta t}[D_t]\right)^+ [\overrightarrow{E_n}] \end{cases} \tag{27}$$

where the averaging operator $[O]$ is given by

$$[O] = \frac{1}{2}\begin{bmatrix} 1 & 0 & 0 & \cdots & 1 \\ 1 & 1 & 0 & \cdots & 0 \\ 0 & 1 & 1 & \cdots & 0 \\ \vdots & \vdots & \ddots & \ddots & \vdots \\ 0 & 0 & \cdots & 1 & 1 \end{bmatrix}_{N \times N} \tag{28}$$

Note that the sample timing of electric field is identical to that of magnetic field in the second equation of (27). In the w domain, we have

$$\begin{cases} \nabla \times \vec{e}_k = +iw_k \mu \; \vec{h}_k \\ \nabla \times \vec{h}_k = (o_k - iw_k \epsilon) \; \vec{e}_k \end{cases} \tag{29}$$

where

$$o_k = \sigma \cos\left[\frac{\pi}{N}(k-1)\right] SGN(k) \tag{30}$$

$o_k$ is symmetric with respect to $k$ (i.e., $o_k = o_{N+2-k}$) and keeps positive over the range of index $k$. Otherwise, a negative $o_k$ will bring about gain rather than loss—obviously it is in conflict with the lossy presumption.

## Supplementary Note 4: w-domain Energy Conservation

We define the averaging Poynting vector of EM periodic sequence as

$$\vec{S}_{av} = \frac{1}{N}\sum_{n=1}^{N} \vec{E}\,[n] \times \left(\frac{\vec{H}[n-1] + \vec{H}[n]}{2}\right) \tag{31}$$

Substituting the inverse transforms of (18) and (19) to (31) under the conjugate symmetric condition, we have

$$\vec{S}_{av} = \frac{1}{N} \sum_{k=1}^{\lceil \frac{N+1}{2} \rceil} \delta_k \, Re(\vec{e}_k \times \vec{h}_k^*) \cos\left[\frac{\pi}{N}(k-1)\right] \tag{32}$$

For odd $N$,

$$\delta_k = \begin{cases} 1, & k = 1 \\ 2, & k = 2,3,\ldots,\frac{N+1}{2} \end{cases} \tag{33}$$

For even $N$,

$$\delta_k = \begin{cases} 1, & k = 1 \ \& \ \frac{N}{2}+1 \\ 2, & k = 2,3,\ldots,\frac{N}{2} \end{cases} \tag{34}$$

Subsequently, the w-domain complex Poynting vector can be represented as

$$\vec{s} = \frac{1}{N} \sum_{k=1}^{\lceil \frac{N+1}{2} \rceil} \delta_k \left(\vec{e}_k \times \vec{h}_k^*\right) \cos\left[\frac{\pi}{N}(k-1)\right] \tag{35}$$

Inserting (29) to (35), we have

$$-\nabla \cdot \vec{s} = i[p_e - p_m] + p_j \tag{36}$$

where

$$p_e = \frac{1}{N} \sum_{k=1}^{\lceil \frac{N+1}{2} \rceil} \delta_k \, w_k \epsilon |\vec{e}_k|^2 \cos\left[\frac{\pi}{N}(k-1)\right] \tag{37}$$

$$p_h = \frac{1}{N} \sum_{k=1}^{\lceil \frac{N+1}{2} \rceil} \delta_k \, w_k \mu |\vec{h}_k|^2 \cos\left[\frac{\pi}{N}(k-1)\right] \tag{38}$$

$$p_j = \frac{1}{N} \sum_{k=1}^{\lceil \frac{N+1}{2} \rceil} \delta_k \, o_k |\vec{e}_k|^2 \cos\left[\frac{\pi}{N}(k-1)\right] \tag{39}$$

$p_e$, $p_h$ and $p_j$ denote the complex power density of electric field, magnetic field and Ohm loss, respectively. (36) describes and substantiates the conservation of energy in the w domain.

## Supplementary Note 5: Recurrent Higher-Order Difference Operator

Difference scheme plays the pivotal role in developing TPS. However, it incurs numerical error due to the approximation of time-partial derivatives. Consequently, it is necessary to analyze this deviation so as to improve the corresponding numerical accuracy.

First, we rewrite (25) as

$$w_k = \omega_k\, Sa\left[\frac{\pi}{N}(k-1)\right], \quad k = 1, \cdots, \frac{N+1}{2} \tag{40}$$

where

$$\omega_k = 2\pi f_k = \frac{2\pi(k-1)}{T} \tag{41}$$

$\omega_k$ denotes the circular frequency that corresponds to the $k^{th}$ index, and factor $Sa[\pi(k-1)/N]$ depicts the numerical approximation which is related to not only the periodic length $N$ but also the spectrum index $k$. For the sake of comparison, we define the normalized bandwidth (NBW) as

$$NBW = \frac{2(k_{max}-1)}{N} \tag{42}$$

As $N$ tends towards infinity and with a fixed value of $k$, the quantized spectrum $w_k$ approaches the circular frequency $\omega_k$. The discrepancy between the two is due to the approximation of the time-partial derivative through difference schemes. Therefore, it is important to carefully select a difference scheme as it directly impacts the numerical accuracy. Fortunately, by introducing the recurrent difference scheme [19], we can improve the accuracy without changing the transform operator (15) and (16).

According to Taylor series expansion, the first-order difference scheme can be expressed as

$$\frac{\psi_{i+1} - \psi_i}{2} = \left[\frac{\frac{\Delta t}{2}}{1!}\frac{\partial}{\partial t} + \frac{\left(\frac{\Delta t}{2}\right)^3}{3!}\frac{\partial^3}{\partial t^3} + \frac{\left(\frac{\Delta t}{2}\right)^5}{5!}\frac{\partial^5}{\partial t^5} + \cdots\right]\psi \tag{43}$$

Based on (43), we can build up the recurrent difference scheme:

$$\frac{\partial}{\partial t} = \frac{D}{\Delta t}\left\{1 + \frac{\left(\frac{\Delta t}{2}\right)^2}{3!}\frac{\partial^2}{\partial t^2}\left[1 + \frac{3!\left(\frac{\Delta t}{2}\right)^2}{5!}\frac{\partial^2}{\partial t^2}(1+\cdots)\right]\right\}^{-1} \tag{44}$$

where $D$ represents the first-order difference scheme $\{-1,1\}$. Note that the first-order difference scheme is explicit whereas the expression of recurrent difference operator is implicit. Subsequently, we use the former one to establish our theoretical model and take the later one for practical computation. Inserting (14) to (44), we have

$$w_k^{(1)} = \frac{w_k}{Sa(w_k/2)} \tag{45}$$

If we denote $w_k$ as $w_k^{(0)}$, the quantized spectrum after $q$ times of recurrent process can be expressed as

$$w_k^{(q)} = \frac{w_k^{(0)}}{Sa(w_k^{(q-1)}/2)} \tag{46}$$

Obviously, given $q \to \infty$, $\omega_k$ is the limitation of $w_k^{(q)}$.

## Supplementary Note 6: Definition of Periodic Excitation

In the w domain, the computational scale depends on the maximum index $k_{max}$ of a given excitation. The general process of obtaining $k_{max}$ is listed as follows:

1. Setting waveshape parameters of a specific waveform in the continuous-time domain, such as (root) raised-cosine, trapezoidal and Gaussian waveform.

2. Cutting off the corresponding spectrum in the continuous-frequency domain, denoted by $f_c$.

3. Deriving $k_{max}$ numerically by solving the optimization problem $\text{argmin}_k\{w_k^{(q)} - 2\pi f_c, w_k^{(q)} \leq 2\pi f_c\}$.

For the sake of analysis, we equidistantly divide period $T$ to $K$ segments, i.e., $T = KT_s$, where $T_s = N_s \Delta t$. As a result, we have $N = KN_s$. In turn, the above optimization problem is equivalent to

$$\text{argmin}_k \left\{ \frac{N_s \sin\left[\frac{\pi(k-1)}{N}\right]}{\pi Sa\left(\frac{w_k^{(q-1)}}{2}\right)} - \frac{f_c}{F_s}, \frac{N_s \sin\left[\frac{\pi}{N}(k-1)\right]}{\pi Sa\left(\frac{w_k^{(q-1)}}{2}\right)} \leq \frac{f_c}{F_s} \right\} \tag{47}$$

where $F_s = 1/T_s$. Note that item $f_c/F_s$ in (47) are dimensionless, so $k_{max}$ is not related to the bandwidth of original signals. This means that time-periodic electromagnetic waves with different bandwidths correspond to the same periodic sequence as long as they have identical waveshape parameters.

Furthermore, considering the ideal scenario that $q \to \infty$, then $k_{max}$ can be explicitly expressed as

$$k_{max} = \left\lfloor \left(\frac{f_c}{F_s}\right)K \right\rfloor \tag{48}$$

where $\lfloor \cdot \rfloor$ denotes the floor-integer operator. Since $f_c/F_s$ is only decided by time-domain waveshape parameters and transform-domain cut-off criterion, we rewrite (48) to

$$k_{max} = \lfloor g(\beta_1, \beta_2, \cdots, \beta_S)K \rfloor \tag{49}$$

Here, $g(\cdot)$ is the function of waveshape and cut-off parameters $\{\beta_s\}_{s=1}^{S}$. A different waveform will result in a different scale of computation. As illustration, $g(\cdot)$ that correspond to raised-cosine, trapezoidal and Gaussian periodic sequences are respectively written as

$$g(\alpha) = \frac{1+\alpha}{2} \tag{50}$$

$$g(r) = \frac{1}{r} \tag{51}$$

and

$$g(tp, fp) = \frac{2}{\pi}\sqrt{ln(tp)ln(fp)} \tag{52}$$

For raised-cosine case (50), $\alpha$ denotes the roll-off coefficient, within the range $[0,1]$. For trapezoidal case (51), $r$ represents the rise-time ratio. For Gaussian case (52), $tp$ and $fp$ depict the truncation positions in time and frequency domain, respectively. Alternatively, we can use time–bandwidth product $BT$ instead of $tp$ to depict the waveshape parameter, i.e.,

$$tp = exp\left[-\left(\pi BT/\sqrt{2ln2}\right)^2\right] \tag{53}$$

## Supplementary Note 7: Analysis of Numerical Dispersion

The numerical dispersion is related to both temporal and spatial discretization. To analyze their effects to the dispersion, we investigate the dispersion relation in a 2D-problem space. Assuming that an infinite plane is filled with lossless, isotropic and homogeneous media, then a w-domain plane wave can be defined as

$$\vec{e_k} = exp[i(\kappa_x x + \kappa_y y)] \tag{54}$$

Partitioning the plane by triangular mesh, we insert (54) to (24) and replacing the infinite boundary condition with Floquet periodic boundary condition[@monk1994dispersion], we can derive the dispersion relation as

$$\kappa_{k_{max}}^2 = \kappa_x^2 + \kappa_y^2 + \frac{\pi^4 \epsilon\mu}{12\Delta t^2} NBW^4 + (\alpha\kappa_1^4 + \beta\kappa_x^3\kappa_y + \gamma\kappa_x^2\kappa_y^2 + \delta\kappa_x\kappa_y^3 + \theta\kappa_y^4) + O(\kappa^4) \tag{55}$$

where $\kappa_k = -\kappa_{N+2-k} = w_k\sqrt{\epsilon\mu}$. $\alpha, \beta, \gamma, \delta$ and $\theta$ represent the geometric coefficients which are decided by a given type of edge element. (55) indicates that the $4^{th}$-order numerical error corresponds to both temporal and spatial discretization. Generally speaking, the spatial discretization can be deemed as a minor contributor given an appropriate mesh (not too coarse). Note that the contribution of spatial discretization will reduce to $6^{th}$-order if the grid is constituted by equilateral triangles. To assess the dispersion level, we define the relative error of propagation constant as

$$p = \frac{|\kappa_{k_{max}} - \omega_{k_{max}}\sqrt{\epsilon\mu}|}{\omega_{k_{max}}\sqrt{\epsilon\mu}} \times 100\% \tag{56}$$

For illustration, we test the relative error of rectangular waveguide (RWG), which also has analytical solution in w domain. Assuming that waveguide is infinitely long in $\hat{z}$ direction (lateral dimension $a \times b$), the wave equation can be expressed as

$$\begin{cases} \nabla^2 \vec{e_k} + \kappa_k^2 \vec{e_k} = 0 \\ \nabla^2 \vec{h_k} + \kappa_k^2 \vec{h_k} = 0 \end{cases} \tag{57}$$

where the w-domain wavenumber is given by

$$\kappa_k = -\kappa_{N+2-k} = w_k\sqrt{\epsilon\mu} \quad, k = 1,\ldots,[\frac{N+1}{2}] \tag{58}$$

In turn, it allows us to use the sign ahead of $\kappa_k$ to discriminate the propagating direction. Similar to frequency domain, the field components in Cartesian coordinate are subjective to

$$\begin{bmatrix} e_{x_k} \\ e_{y_k} \\ h_{x_k} \\ h_{y_k} \end{bmatrix} = \frac{1}{\kappa_c^2} \begin{bmatrix} \frac{\partial}{\partial z} & 0 & 0 & iw_k\mu \\ 0 & \frac{\partial}{\partial z} & -iw_k\mu & 0 \\ 0 & -iw_k\epsilon & \frac{\partial}{\partial z} & 0 \\ iw_k\epsilon & 0 & 0 & \frac{\partial}{\partial z} \end{bmatrix} \begin{bmatrix} \frac{\partial e_{z_k}}{\partial x} \\ \frac{\partial e_{z_k}}{\partial y} \\ \frac{\partial h_{z_k}}{\partial x} \\ \frac{\partial h_{z_k}}{\partial y} \end{bmatrix} \tag{59}$$

Correspondingly, the TE$_{mlk}$ mode of RWG can be expressed as

$$
\begin{aligned}
h_{z_k} &= h_{ml}^k \cdot 1 \cdot \cos(\kappa_x^m x)\cos(\kappa_y^l y)exp(i\kappa_{z_k} z) \\
e_{x_k} &= h_{ml}^k \cdot \frac{-i\mu\kappa_y^l w_k}{\kappa_c^2} \cdot \cos(\kappa_x^m x)\sin(\kappa_y^l y)exp(i\kappa_{z_k} z) \\
e_{y_k} &= h_{ml}^k \cdot \frac{i\mu\kappa_x^m w_k}{\kappa_c^2} \cdot \sin(\kappa_x^m x)\cos(\kappa_y^l y)exp(i\kappa_{z_k} z) \\
h_{x_k} &= h_{ml}^k \cdot \frac{-i\kappa_{z_k}}{\kappa_c^2} \cdot \sin(\kappa_x^m x)\cos(\kappa_y^l y)exp(i\kappa_{z_k} z) \\
h_{y_k} &= h_{ml}^k \cdot \frac{-i\kappa_{z_k}}{\kappa_c^2} \cdot \cos(\kappa_x^m x)\sin(\kappa_y^l y)exp(i\kappa_{z_k} z)
\end{aligned}
\quad (60)
$$

Where $h_{ml}^k$ represents the magnetic field intensity of the input periodic sequence, $m$ and $l$ denote the modal indices of $\hat{x}$ and $\hat{y}$ directions, respectively; $\kappa_x^m = \frac{m\pi}{a}$, $\kappa_y^l = \frac{l\pi}{b}$, $\kappa_c = \sqrt{(\kappa_x^m)^2 + (\kappa_y^l)^2}$ and $\kappa_{z_k}$ denotes the w-domain dispersion of RWG:

$$\kappa_{z_k} = \sqrt{\kappa_k^2 - \left(\frac{m\pi}{a}\right)^2 - \left(\frac{l\pi}{b}\right)^2} \quad (61)$$

Here $a$ and $b$ represent the width and height of RWG respectively, and $(m, l)$ stands for the spatial-modal index pair. Since (61) is related to the index $k$, we stipulate that the complete set of $\{TE_{10k}\}$ $(k = 1, \cdots, k_{max})$ represents the 'TE$_{10}$ mode' of RWG.

If the magnitude of $\kappa_{z_k}$ greatly exceeds that of the third item on the right hand side of (55), then (56) can be approximated as follows:

$$p = \frac{\pi^2}{24} NBW^2 \quad (62)$$

(62) preserves a good consistency with the theoretical result within a considerable span of NBW. Similarly, the approximation of (56) corresponding to the first- and second-order recurrence can be expressed by

$$p = \frac{\pi^4}{288} NBW^4 \quad (63)$$

and

$$p = \frac{\pi^6}{3456} NBW^6 \quad (64)$$

, respectively. Given a fixed accuracy level, the qualified span of NBW is extended with the increase of recurrent order, leading to a well-restrained dispersion level.

## Supplementary Note 8: w-domain S-Parameters

Note that the aforementioned theory only involves the temporal part of Maxwell's curl equations—we should introduce the spatial-part process in the w domain so as to solve general EM problems. Without loss of generality, a finite element method (FEM) is chosen as the w-domain solver, in which EM fields are represented by first-type Nédélec edge elements [21]. In this manuscript, we focus on closed region scenario.

Assuming that the close-region problem only involves two types of boundary conditions, namely 'PEC' and 'wave port'. For each index k, the matrices equation can be expressed as [22]

$$\{[P^e] - \kappa_k^2[Q^e] + iw_k\mu_0 \sum_{p=1}^{P}\sum_{m=1}^{\infty}[b_{mk}^p][b_{mk}^p]^T\}[e_k]$$
$$= 2iw_k\mu_0 \sum_{p=1}^{P}\sum_{m=1}^{\infty}[b_{mk}^p] \tag{63}$$

where $[P^e]$, $[Q^e]$ and $[b_{mk}^p]$ are given by

$$(P^e)_{ij} = \iiint_{V_e} \left(\nabla \times \vec{N}_i^e \cdot \bar{\mu}^{-1} \cdot \nabla \times \vec{N}_j^e\right) dV \tag{64}$$

$$(Q^e)_{ij} = \iiint_{V_e} \left(\vec{N}_i^e \cdot \bar{\epsilon} \cdot \vec{N}_j^e\right) dV \tag{65}$$

$$(b_{mk}^p)_i = \iint_{S_e^p} \left(\vec{N}_i^e \times \vec{h}_{mk}^p\right) \cdot \hat{n} \, dS \tag{66}$$

, respectively; $\bar{\epsilon}$ ($\bar{\mu}$) is the tensor form of relative permittivity (permeability), subscript $i$ ($j$) denote the global index of the vector basis, $\vec{N}_{i,j}^e$ represents the vector finite element basis of the tetrahedral (triangular) element $V_e$ ($S_e^p$), $\vec{h}_{mk}^p$ is the normalized magnetic modal field, the superscript $p$ denotes the $p^{th}$ port ($P$ ports total) and the superscript $m$ represents the spatial index of the modal field.

Once (65) is solved, the w-domain scattering parameters can be defined as

$$s_{mk}^{pp_0} = [b_{mk}^p]^T \cdot [e_k] - \delta_{pp_0} \tag{67}$$

Here, the periodic excitation is only assigned to the $p_0^{th}$ port. If $p = p_0$, $\delta_{pp_0}$ equals to one otherwise zero. In addition, the eigenvalue problem on each port surface is solved by modifying the equation (27) [23].

After obtaining the w-domain S-parameters, it is easily to derive the output response for arbitrary port. The overall process can be summarized as

$$\tilde{q}[n] = \tilde{p}[n] \odot \tilde{h}[n] \tag{68}$$

where $\tilde{p}[n]$, $\tilde{h}[n]$ and $\tilde{q}[n]$ represent the input excitation, the impulse response of the close-region and the corresponding output response, respectively; the tilde sign indicates that the sequence is periodic type and $\odot$ represents the $N$-points circular convolution. Note that the response is only related to the sequence—all spatial variables have been eliminated by the surface integration on port.